\documentclass[a4paper]{jpconf}
\usepackage{graphicx}
\graphicspath{{figs_pdf/}}
\usepackage{color}
\usepackage{xcolor}
\definecolor{rev1}{rgb}{0,0,0}
\usepackage{amsmath,amssymb,amsfonts}
\usepackage{hyperref}
\begin{document}
\title{Hybrid analysis and modeling for next generation of digital twins}

\author{Suraj Pawar$^1$, Shady E. Ahmed$^1$, Omer San$^1$, Adil Rasheed$^2$}

\address{$^1$School of Mechanical \& Aerospace Engineering, Oklahoma State University, Stillwater, OK 74078, USA. \\
$^2$Department of Engineering Cybernetics, Norwegian University of Science and Technology, 7465 Trondheim, Norway.}

\ead{supawar@okstate.edu, shady.ahmed@okstate.edu, osan@okstate.edu, adil.rasheed@ntnu.no}

\begin{abstract}
The physics-based modeling has been the workhorse for many decades in many scientific and engineering applications ranging from wind power, weather forecasting, and aircraft design. Recently, data-driven models are increasingly becoming popular in many branches of science and engineering due to their non-intrusive nature and online learning capability. Despite the robust performance of data-driven models, they are faced with challenges of poor generalizability and difficulty in interpretation. These challenges have encouraged the integration of physics-based models with data-driven models, herein denoted hybrid analysis and modeling (HAM). We propose two different frameworks under the HAM paradigm for applications relevant to wind energy in order to bring the physical realism within emerging digital twin technologies. The physics-guided machine learning (PGML) framework reduces the uncertainty of neural network predictions by embedding physics-based features from a simplified model at intermediate layers and its performance is demonstrated for the aerodynamic force prediction task. Our results show that the proposed PGML framework achieves approximately 75\% reduction in uncertainty for smaller angle of attacks. The interface learning (IL) framework illustrates how different solvers can be coupled to produce a multi-fidelity model and is successfully applied for the Boussinesq equations that govern a broad class of transport processes. The IL approach paves the way for seamless integration of multi-scale, multi-physics and multi-fidelity models ($M^{3}$ models).
\end{abstract}

\section{Introduction}
Some of the major challenges in realizing the potential of wind energy to meet the global electricity demand are the need for a deeper understanding of the physics of the atmospheric flow, science, and engineering of these large dynamic rotating machines and synergistic optimization and control of fleets of wind farms within the electricity grid \cite{veers2019grand}. The decades of research and development in fluid dynamics, systems engineering, manufacturing processes, material discovery can now be complemented with unprecedented amounts of data generated from in-situ measurements, lab experiments, and numerical simulations to tackle these challenges. The combination of physics-based and data-driven models is increasing in every branch of science leading to the hybrid analysis and modeling (HAM) approach for many scientific applications.   

\begin{figure*}[htbp]
\centering
\includegraphics[width=0.95\textwidth]{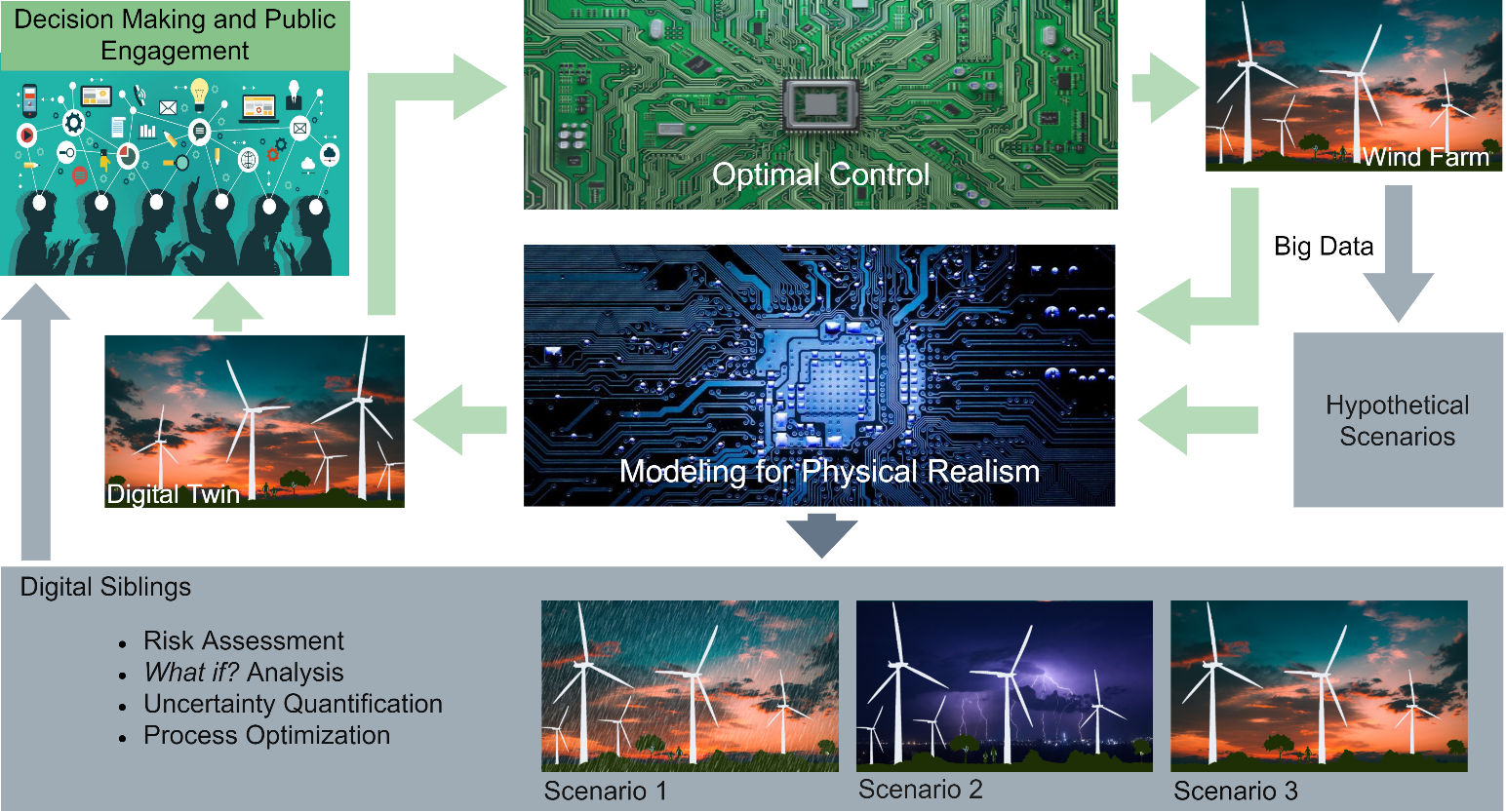}
\caption{Digital twin of the wind farm can allow us to evaluate different scenarios using hybrid models. The big data collected from IoT sensors can be continuously assimilated to correct hybrid models and improve the state estimation. }
\label{fig:dt}
\end{figure*}

The HAM paradigm can be applied to a variety of tasks related to wind research, such as digital twinning for the optimization and real-time control of wind farms. A digital twin is defined as a virtual replica of a physical system enabled through data collected from sensors and simulations in real-time to solve problems such as control, optimization, monitoring, and improved decision-making \cite{rasheed2020digital}. The advancement in multiphysics solver, computational infrastructure, big data, and artificial intelligence has allowed us to make a digital replica of large physical systems such as wind farms \cite{DT_wind_farm}. The digital twin lets us examine \emph{what if} scenarios, evaluate the system's response, and select the corresponding mitigation strategies. Figure~\ref{fig:dt} shows the typical digital twin framework for a wind farm. The digital twin of wind farms can be useful for different purposes such as optimal control of wind turbines to achieve the maximum performance, routine maintenance of the equipment, accurate forecast of the power production, wind farm optimization \cite{barter2020systems}, and improved decision-making.            

The success of the digital twins depends upon the type of approach that we employ for modeling the system. The ability of the HAM to combine the generalizability of physics-based approaches and the automatic pattern-identification feature of data-driven approaches makes it an attractive choice to model virtual replica of physical systems in digital twins. For example, there are several turbine wake models that capture the flow in the wake of a single turbine and these models predict sufficiently accurate aerodynamics of a single turbine \cite{bastankhah2014new,shapiro2018modelling}. However, these models are not enough to capture the flows in wind farm due to wake superposition, complex terrains, deep array effects, and neglected physics \cite{meneveau2019big,politis2012modeling,schlez2009new}. The observational data collected from a variety of sensors can be assumed to comprise of these complex flow interactions and the manifestation of all physical processes. Therefore, the hybrid model will aid in the robustness of the digital twin rather than a pure physics-based or pure data-driven approach. 

In this work, we introduce two different frameworks under the umbrella of HAM. The first framework is called physics-guided machine learning (PGML) where information from simplified physics-based models is incorporated within neural network architectures to improve the generalizability of data-driven models \cite{pawar2020physics}. The second framework is our interface learning approach that deals with coupling different solvers and mathematical descriptions using statistical inference tools \cite{pawar2020interface,ahmed2020interface,ahmed2020interfaceHAM}. One of the potential applications of this framework can be to specify the physically accurate inlet boundary condition for the wind farm simulation. For example, it is very common to use a coarse-grid solver to resolve the atmospheric boundary layer flow and employ the fine-grid solver to resolve the flow field around wind turbines in the wind farm. For such problems, it is very important to exchange the information between two solvers that leads to physically consistent boundary conditions, and statistical tools like deep learning can be exploited for this. 

Concisely, bringing physical realism in digital twins will need 
    \begin{itemize}
        \item new modelling approaches that are accurate and certain, generalizable, computationally efficient, and trustworthy,
        \item seamless integration of multi-scale, multi-physics and multi-fidelity models ($M^{3}$ models).
    \end{itemize} 
To this end, we propose two HAM approaches that address the above needs. While the first approach PGML provides a mechanism to guide the learning of a machine learning algorithm using simplified physics, the second approach IL enables coupling of $M^{3}$ models.

\section{Physics-guided machine learning}
A wide variety of problems in wind energy such as aerodynamic performance prediction \cite{zha2007high,legresley2000airfoil} for optimal control, design optimization, uncertainty quantification requires the prediction of the quantity of interest in real-time. The prediction of flow around an airfoil is a high-dimensional, and nonlinear problem that can be solved using high-fidelity methods like computational fluid dynamics (CFD). However, these methods are computationally infeasible for real-time prediction. On one hand, in certain flow regimes, the simplified methods like panel codes come with a non-negligible difference between the actual dynamics and approximate models for real-world problems. On the other hand, the full-order CFD simulations are computationally demanding, thus limiting their use in many inverse modeling methodologies that require a model run to be performed in each iteration. To overcome these challenges, combining CFD models with machine learning to build a non-intrusive surrogate model is gaining widespread popularity \cite{zhang2018application,bhatnagar2019prediction}. One of the main challenges with these non-intrusive models is their predictive capabilities for unseen data and their interpretation. Even though there are methods to predict the uncertainty estimate in the prediction of machine learning models \cite{zhu2019physics,maulik2020probabilistic}, the generalizability of non-intrusive models is not on par with physics-based models. Meanwhile, the simplified models like the Blasius boundary layer model in fluid mechanics are highly generalizable across different conditions. Therefore, it is important to incorporate the domain knowledge into learning, and to this end, we exploit the relevant physics-based features from panel methods through the PGML framework to enhance the generalizability of data-driven surrogate model.  

\begin{figure*}[htbp]
\centering
\vspace{-0.2in}
\includegraphics[width=0.99\textwidth]{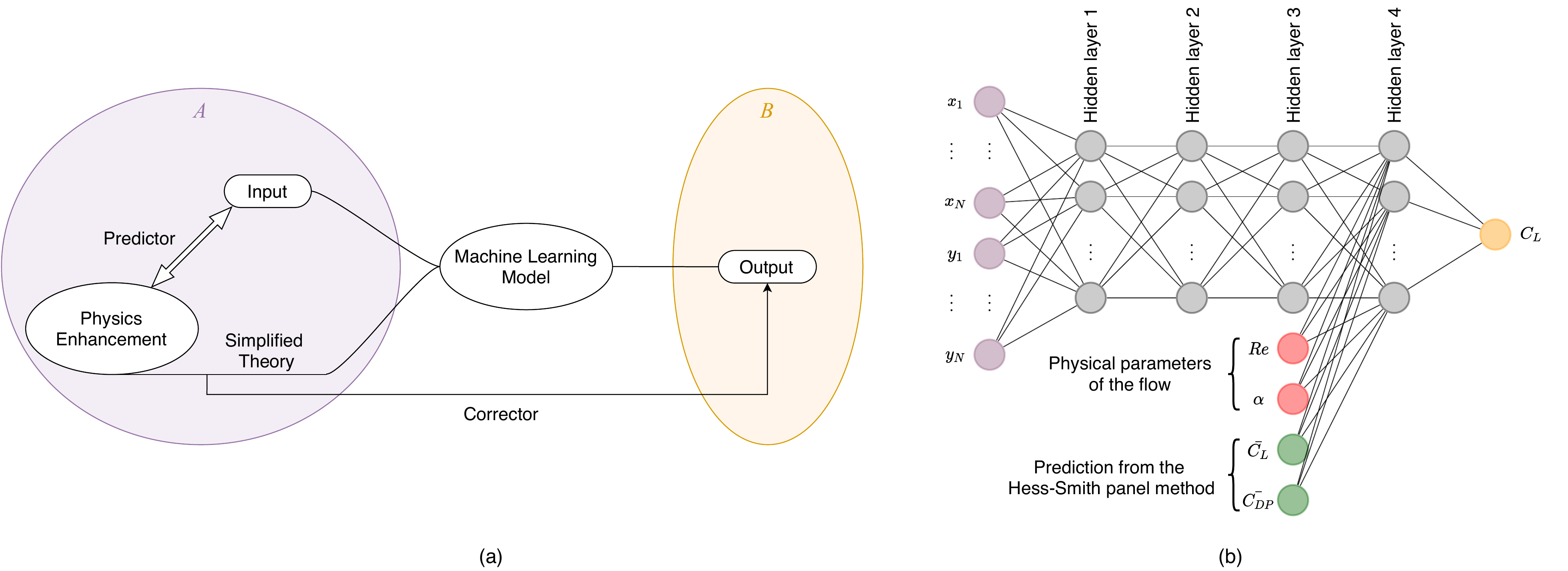}
\caption{Physics guided machine learning (PGML) framework to train a learning engine between processes $A$ and $B$: (a) a conceptual PGML framework which shows a predictor-corrector approach to incorporate physics into machine learning models through embedding simplified theories directly into neural network models, and (b) the representative neural network architecture of the PGML framework used in this study for aerodynamic forces prediction task.}
\label{fig:PGNN}
\end{figure*}

We now introduce different components of the PGML framework depicted in Figure~\ref{fig:PGNN}(a). In supervised machine learning, the input vector $\mathbf{x} \in \mathbb{R}^m$ is fed to the machine learning model (for example, the neural network in our case), and the mapping from the input vector to the output vector $\mathbf{y} \in \mathbb{R}^n$ is learned through training. The neural network is trained to learn the function $F_\theta$, parameterized by $\theta$, that includes the weights and biases of each neuron. The parameters of the neural network are optimized using the backpropagation algorithm to minimize the cost function. Usually, for the regression problems, the cost function is the mean squared error between true and predicted output, i.e., $C(\mathbf{x},\theta)=||\mathbf{y} -F_{\theta}(\mathbf{x})||_2$. In the PGML framework, the neural network is augmented with the output of the simplified physics-based model. The features extracted from simplified physics-based models are embedded into hidden layers along with latent variable. This is in contrast to admitting physics-based features at the input layer in conventional neural network architectures, which might lead to underestimation of the effect of such physical information, especially for high dimensional systems. For example, imagine stacking a few parameters to an input vector of a dimension of $O(10^3-10^6)$. It is highly probable that the learning algorithm overlooks the effect of such prior knowledge in the minimization algorithm, and a modification of the cost function becomes necessary. In the PGML, a proper latent space is first identified and the information from the physics-based model aids the neural network in constraining its output to a manifold of physically realizable models.

The training data for the neural network is generated using a series of numerical simulations performed in XFOIL \cite{drela1989xfoil}. We highlight here that the the neural network can also be trained using the data gathered from CFD simulations or wind-tunnel experiments. The lift coefficient data were obtained for different Reynolds numbers $Re$ between $1 \times 10^6$ and $4 \times 10^6$ and several angles of attacks $\alpha$ in the range of $-20$ to $+20$. A total of 168 sets of two-dimensional airfoil geometry were generated from NACA 4-digit, NACA210, NACA220, and NACA250 series for training the neural network. Each airfoil is represented by 201 points. The maximum thickness of all airfoils in the training dataset was between 6\% to 18\% of the chord length. We use the NACA23012 and NACA23024 airfoil geometry as the test dataset to evaluate the predictive capability of the trained neural network. The simplified model used to generate the physics-based feature corresponds to the Hess-Smith panel method \cite{hess1990panel} based on potential flow theory. We note here that our testing airfoils are selected not only from a different NACA230 series (i.e., not used in the training dataset), but also the maximum thickness of 24\% is well beyond the thickness ratio limit included in the training dataset.   

\begin{figure}[htbp]
\centering
\vspace{-0.2in}
\includegraphics[width=0.75\textwidth]{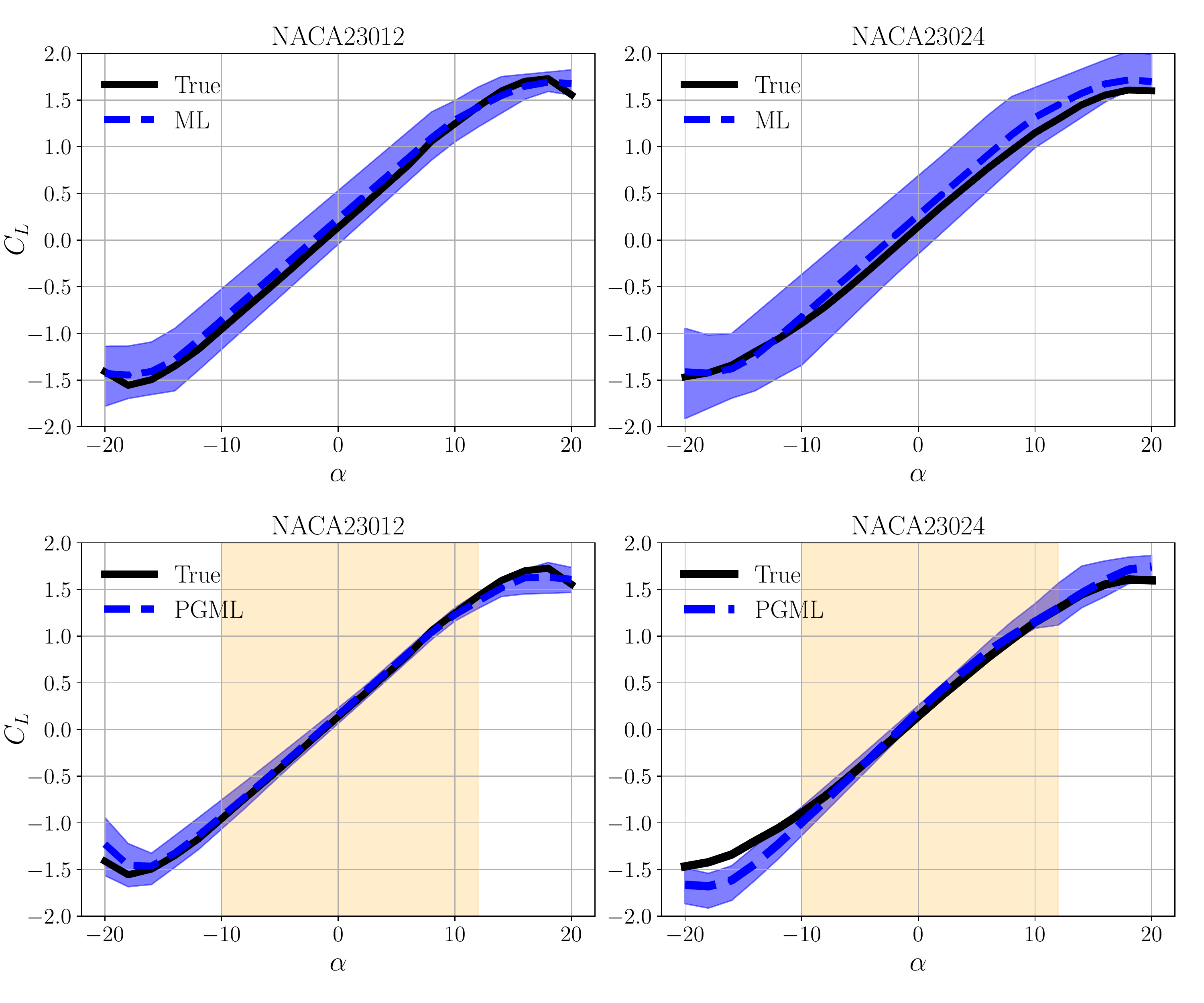}
\vspace{-0.2in}
\caption{Actual versus predicted lift coefficient ($C_L$) for NACA23012 and NACA23024 airfoils at $Re=3 \times 10^6$ using ML and PGML framework. The dashed blue curve represent the average of the predicted lift coefficient by all data-driven models (i.e., testing runs with different initialization seeds).} 
\label{fig:airfoil_nn}
\end{figure}

The adopted neural network architecture has four hidden layers with 20 neurons in each hidden layer. The physical parameters, i.e., the Reynolds number and the angle of attack are concatenated at the third hidden layer along with the latent variables at that layer. In the PGML model, we augment the latent variables at the third layer with the lift coefficient and the pressure drag coefficient predicted by the panel method along with physical parameters of the flow (i.e., the Reynolds number and angle of attack). Therefore, the third layer of the neural network in the PGML framework has 24 latent variables. The representative neural network for the PGML framework to predict the aerodynamic forces on an airfoil is displayed in Figure~\ref{fig:PGNN}(b). We utilize an ensemble of neural networks trained using different initialization to predict the epistemic uncertainty \cite{tibshirani1996comparison,lakshminarayanan2017simple}. The weights and biases of each model are initialized using the Glorot uniform initializer and different random seed numbers are used to ensure that different values of weights and biases are assigned for each model. The ensemble of all these models indicates the model uncertainty estimate of the predicted lift coefficient. Figure~\ref{fig:airfoil_nn} shows the actual and predicted lift coefficient for the NACA23012 and NACA23024 airfoil geometry. The reference \emph{True} performance is obtained by XFOIL. The ML corresponds to a simple feed-forward neural network that uses the airfoil $x$ and $y$ coordinates as the input features, and the physical parameters of the flow are concatenated at the third hidden layer along with the latent variables at that layer.

As shown in Figure~\ref{fig:airfoil_nn}, we can see that the uncertainty in the prediction of the lift coefficient by the PGML model is considerably less than the ML model for both NACA23012 and NACA23024 airfoils. The proposed PGML framework provides significantly more accurate predictions with uncertainty reduced approximately by 75\% especially for the angle of attacks between -10 and +12 degrees. This further illustrates the viability of the proposed PGML framework, since the physics embedding considered here employs constant source panels and a single vortex to approximate the potential flow around the airfoil. We can also notice that the uncertainty is higher for the angle of attacks outside the range of -10 to +12 degrees. This finding is not surprising as the Hess-Smith panel method is a proven method for analysis of inviscid flow over airfoil for the smaller angle of attacks regime. The maximum thickness of an airfoil included in the training dataset is 18\% of the chord length. Therefore, the uncertainty in the prediction of the lift coefficient by the ML model is higher for the NACA23024 airfoil compared to the NACA23012 airfoil. These results clearly show the potential of the PGML framework for building trustworthy models that can enable the digital twin of physical systems.

\section{Interface learning}
\begin{figure}[htbp]
\centering
\includegraphics[width=0.99\textwidth]{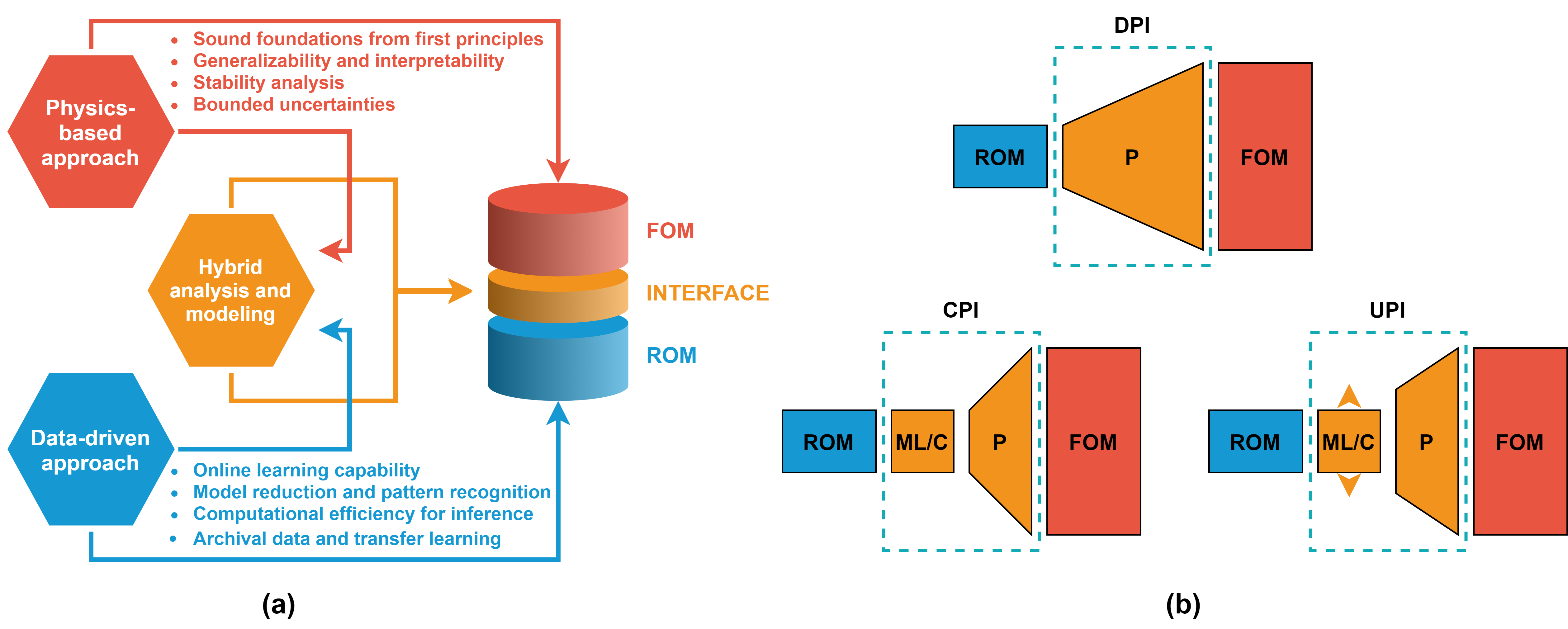}
\vspace{-0.1in}
\caption{Hybrid analysis and modeling (HAM) as a key enabler for ROM-FOM coupling problems toward predictive digital twins: (a) an overview, and (b) proposed ROM-FOM coupling frameworks.}
\vspace{-0.2in}
\label{fig:hamFOMROM}
\end{figure}

The second framework we are introducing under the umbrella of HAM is the interface learning (IL). Multi-scale, multi-physics and multi-fidelity models ($M^{3}$ models) are the main beneficiaries from the IL methodology. Many complex systems relevant to scientific and engineering applications include multiple spatiotemporal scales and comprise a multifidelity problem sharing an interface between various formulations or heterogeneous computational entities. We refer the readers to our previous discussion about the potential of IL approaches, with demonstrations on truncated domains \cite{ahmed2020interface}, and mico-macro scale solvers coupling \cite{pawar2020interface}. In the present study, we are interested in situations where part of the domain, physics, or scales are characterized by repeating coherent structures, and can thus be represented by a reduced order model (ROM) for computational speed-up. In the meantime, a high-fidelity full order model (FOM) is dedicated for the rest of domain/dynamics for accuracy requirements. However, both solvers are coupled and information should be communicated and matched at their \emph{interface}. To this end, we present a robust HAM approach combining a physics-based FOM and a data-driven ROM to form the building blocks of an integrated approach among mixed fidelity descriptions toward predictive digital twin technologies, as depicted in Figure~\ref{fig:hamFOMROM}(a).

\subsection{ROM-FOM coupling framework}
In order to demonstrate the ROM-FOM coupling framework, we consider a coupled system as follows, 
\begin{align}
    \dfrac{\partial u}{\partial t} = f_1(u;\mu_1) + g_1(u,v;\mu_1,\mu_2), \label{eq:u}\\
    \dfrac{\partial v}{\partial t} = f_2(v;\mu_2) + g_2(u,v;\mu_1,\mu_2),
\end{align}
where $u$ and $v$ are the coupled variables and $g_1$ and $g_2$ define this coupling, while $\mu_1$ and $\mu_2$ denote the set of system's parameters. We highlight that the coupled variables might represent the state variables at different regions of the domain (e.g., multi-component systems), different physics (e.g., fluid-structure interactions) and/or different scales within the same domain (e.g., multiscale systems). We suppose that the dynamics of $u$ can be approximated by a ROM while a FOM resolves $v$ and both solvers need to communicate information to satisfy the coupling.

\subsubsection{Reduced order model}
Introducing a spatial discretization to Eq.~(\ref{eq:u}), it can be rewritten in a semi-discrete continuous-time as follows, 
\begin{equation}
    \dfrac{\mathrm{d} \mathbf{u}}{\mathrm{d} t} = \mathcal{F}(\mathbf{u}, \mathbf{v}; \boldsymbol{\mu}) = \mathcal{L}_1 \mathbf{u} + \mathcal{L}_2 \mathbf{v} + \mathcal{N}(\mathbf{u},\mathbf{v}), \label{eq:udis}
\end{equation}
where the boldfaced symbols represent the arrangement of discretized variables in a column vector, $\boldsymbol{\mu}$ defines the system's parameters, and $\mathcal{F}$ is a deterministic operator with linear and nonlinear components $\mathcal{L}$, and $\mathcal{N}$, respectively. We exploit the advances and developments of ROM techniques to build surrogate models to economically resolve portions of domain and/or physics. The ROM solution can thus be used to infer the flow conditions at the interface so that a FOM solver can be efficiently employed for the regions of interest. The standard Galerkin ansatz is applied for the dynamics of $\mathbf{u}$ as $\mathbf{u}(t) \approx \Phi \boldsymbol{\alpha}(t)$, 
where the columns of matrix $\Phi = [\phi_1, \phi_2, \dots, \phi_r]$ form the orthonormal bases of a reduced subspace, and $\boldsymbol{\alpha}$ defines their amplitudes. Proper orthogonal decomposition (POD) is one popular technique to systematically construct $\Phi$ such that the solution manifold preserves as much variance as possible when projected onto the subspace spanned by $\Phi$ \cite{holmes2012turbulence}. By substituting this approximation into Eq.~(\ref{eq:udis}), performing the inner product with $\Phi$, and making use of the quadratic nonlinearity in most fluid flow systems, we get the following,
\begin{equation} 
\dfrac{\mathrm{d} \boldsymbol{\alpha}}{\mathrm{d} t} = \mathbf{\mathfrak{L}} \boldsymbol{\alpha}  +  \boldsymbol{\alpha}^T \mathbf{\mathfrak{N}} \boldsymbol{\alpha} + \mathbf{\mathfrak{C}} , \label{eq:ROM}
\end{equation}
where $\mathbf{\mathfrak{L}}$ and $\mathbf{\mathfrak{N}}$ signify the model coefficients while $\mathbf{\mathfrak{C}}$ defines the contribution of $\mathbf{v}$ into the ROM of $\mathbf{u}$. At the interface, we introduce a long short-term memory network to bridge the low-fidelity descriptions to high-fidelity models in various forms of interfacial error correction or prolongation. An array of interface modeling paradigms are sketched in Figure~\ref{fig:hamFOMROM}(b) and summarized as follows (see \cite{ahmed2020interfaceHAM} for more details),

\begin{enumerate}
    \item \emph{DPI: Direct Prolongation Interface.} The DPI approach provides an estimate of the flow variables at the interface from the ROM solution. Indeed, this prolongation map naturally results from the Galerkin ansatz, without any interference from the ML side. However, it is known that truncated Galerkin ROMs might yield erroneous and even unstable predictions for complex systems \cite{ahmed2020long}. Therefore, the solution from the DPI approach is potentially inaccurate, and a correction needs to be introduced. 
    
    \item \emph{CPI: Correction followed by Prolongation Interface.} The CPI methodology works by introducing the correction in the latent subspace and addresses the deviation in modal coefficients predicted from solving the Galerkin ROM, known as closure error. Specifically, the LSTM for CPI takes the values of modal coefficients acquired from integrating Eq.~(\ref{eq:ROM}) and predicts the discrepancy between these values and their optimal values. We highlight here that the size of the input and output vectors is $O(r)$, independent of the FOM resolution, which offers a potential flexibility dealing with 2D and 3D problems.

    
    \item \emph{UPI: Uplifted Prolongation Interface.} Although the CPI methodology cures the closure error and provides a stabilized solution, it does not address the projection error. Unless a large number of modes are resolved, the projection error can be significant. An uplifting ROM has been proposed \cite{ahmed2020long}, where both closure and projection errors are taken care of. In addition to the closure modeling, the ROM subspace is expanded to recover some of the smaller scales missing in the initial subspace as follows,
    \begin{equation}
        \mathbf{u} \approx \Phi \boldsymbol{\alpha} + \Psi \boldsymbol{\beta},
    \end{equation}
    where $\Psi$ forms orthonormal basis for a $q$-dimensional subspace complementing that spanned by $\Phi$ and with $\boldsymbol{\beta}$ being the corresponding amplitudes. We highlight that the Galerkin ROM equations only solve for $\boldsymbol{\alpha}$ to keep the computational cost as low as possible. Therefore, a complementary model for $\boldsymbol{\beta}$ has to be constructed so that the uplifting approach can be employed. A mapping from the first $r$ modal coefficients to the next $q$ modes is assumed to exist and we exploit the LSTM learning capabilities to infer this map from data. In particular, the UPI architecture is trained to read the Galerkin ROM prediction for the first $r$ modal coefficients as input, and return the true coefficients of the first $r+q$ modes. Thus, it provides a \emph{closure} correction for the first $r$ modes and a \emph{superresolution} effect for the next $q$ modes, simultaneously in a single network. 
\end{enumerate}

\subsection{Demonstration using Boussinesq equations}
The dimensionless form of the 2D incompressible Boussinesq equations can be represented by the following two coupled transport equations in vorticity-streamfunction formulation,
\begin{align}
\dfrac{\partial \omega}{\partial t} + J(\omega,\psi) &= \dfrac{1}{\text{Re}}\nabla^2 \omega + \text{Ri} \dfrac{\partial \theta}{\partial x}, \label{eq:Bouss1} \\
\dfrac{\partial \theta}{\partial t} + J(\theta,\psi) &= \dfrac{1}{\text{Re} \text{Pr}} \nabla^2 \theta, \label{eq:Bouss2}
\end{align}
where $\omega$, $\psi$ and $\theta$ denote the vorticity, streamfunction and temperature fields, respectively. We utilize the 2D Boussinesq equation to illustrate the ROM-FOM coupling in multi-fidelity environments. In particular, we suppose that we are more interested in the temperature field predictions. Thus, we dedicate a FOM solver for Eq.~(\ref{eq:Bouss2}). However, the solution of this equation requires evaluating the streamfunction field at each time step. The kinematic relationship between vorticity and streamfunction is given by the Poisson equation (i.e, $\nabla^2 \psi = -\omega$), the solution of which consumes significant amount of time and computational resources and is considered the bottleneck for most incompressible flow solvers. Therefore, we consider a ROM solver for the voriticity dynamics and FOM solver for the temperature field. 

For demonstration, we explore the lock-exchange problem, defined by two fluids of different temperatures, in a rectangular domain $(x,y) \in [0,8] \times [0,1]$ with a vertical barrier dividing the domain at $x=4$, keeping the temperature of the left half at $1.5$ and temperature of the right half at $1$. Initially, the flow is at rest, with uniform temperatures at the right and left regions, and the barrier is removed at $t=0$. Reynolds number of $\text{Re} = 10^4$, Richardson number of $\text{Ri} = 4$, and Prandtl number of $\text{Pr}=1$ are set and a Cartesian grid of $4096\times512$ with a timestep of $\Delta t=5\times10^{-4}$ are used for the FOM simulations.

\begin{figure}[!ht]
    \centering
    \vspace{-0.2in}
    \includegraphics[width=0.62\textwidth]{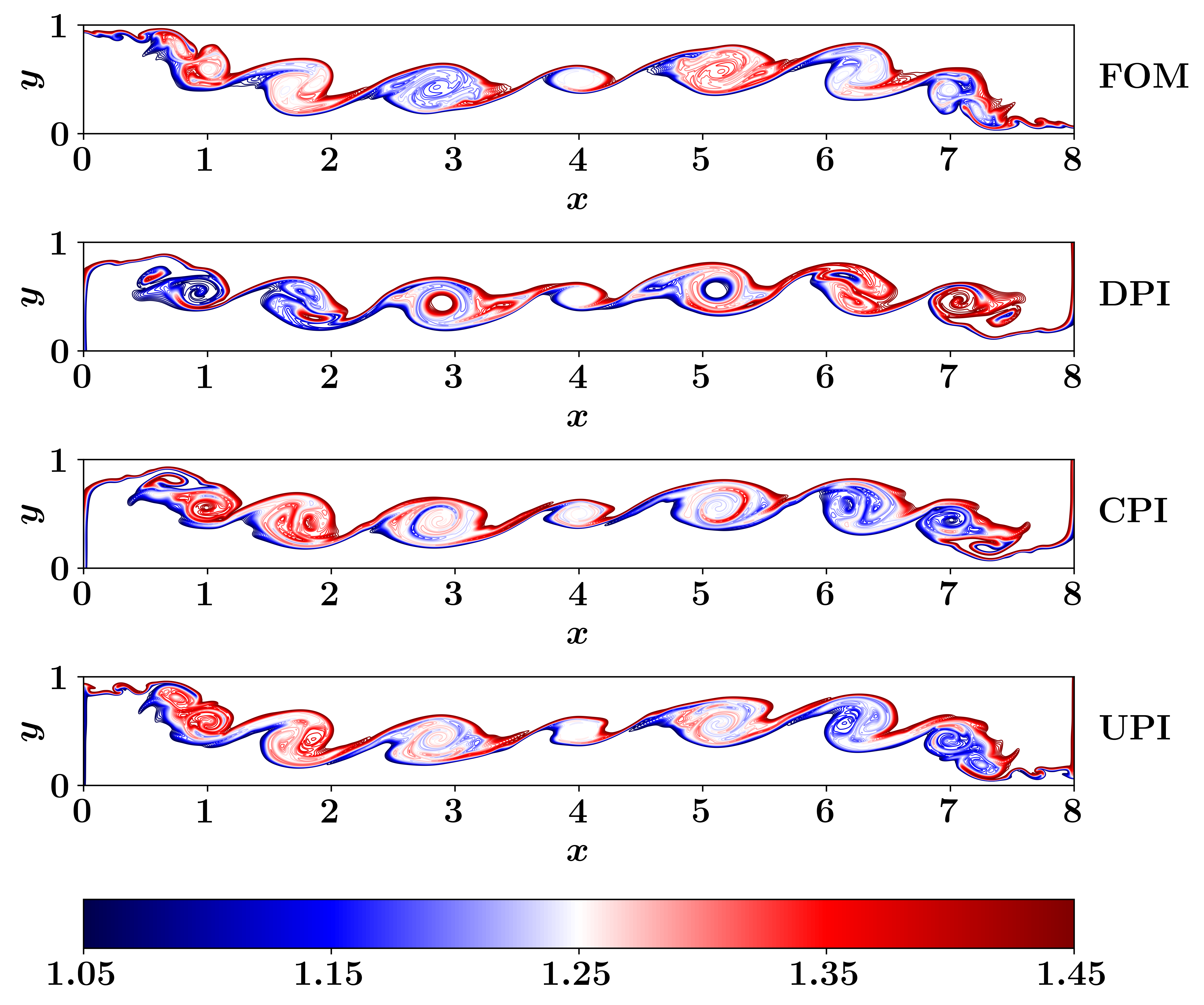}
    \vspace{-0.1in}
    \caption{Final temperature fields as obtained from different ROM-FOM coupling approaches, compared to the FOM solution.}
    \label{fig:BScoup}
\end{figure}

A two-layer LSTM with 20 hidden units in each LSTM cell constitutes our ML architecture. We store $800$ time snapshots for POD basis construction and we retain $r=8$ modes for the Galerkin ROM solver. We also utilize the dataset of the stored $800$ snapshots for LSTM training and validation. During the testing phase, the trained neural networks are deployed at each and every timestep. This corresponds to the deployment of the presented approached $16000$ times. Figure~\ref{fig:BScoup} shows the predictions of the temperature field at final time (i.e., $t=8$) computed from DPI, CPI, and UPI approaches compared to the FOM field. We emphasize that the ROM-FOM coupling results correspond to the solution of the vorticity equation with a ROM solver, which feeds the FOM solver with streamfunction to solve the temperature equation only as opposed to the FOM results which comes from the solution of both equations using a fully FOM simulation. Although the CPI results are better than those of DPI, we can observe that the fine details of the flow field are not accurately captured. On the other hand, the implementation of the UPI approach with $r=8$ and $q=8$ recovers an increased amount of the fine flow structures that are not well-represented by the first $r=8$ modes.

\section{Concluding remarks}

The data-driven methods are increasingly being applied in many branches of science and engineering due to their success in automatic pattern-identification using the data collected from sensors and numerical simulations. Even though they offer an alternative to physics-based modeling derived from phenomenological arguments, they are usually black box in nature and lack generalizability. The hybrid analysis and modeling (HAM) is a newly emerging paradigm that combines physics-based and data-driven modeling to deliver robust and generalizable models that can enable the digital twins of large scale physical systems.    

The major contributions of this work towards HAM paradigm are
\begin{itemize}
    \item A novel deep neural network architecture that makes it possible to inject physics during the training process. This resulted in a significant reduction of uncertainty.
    \item An interface learning technique that makes seamless coupling of multi fidelity models possible.
\end{itemize}

The physics-guided machine learning (PGML) framework enhances the generalizability of the neural network based surrogate model and its performance is illustrated for real time aerodynamic performance prediction task. The interface learning (IL) framework allows integration of two different models seamlessly to build multi-fidelity models that are orders of magnitude faster than full order physics-based models. Although PGML approach was demonstrated to work with deep neural network, it can easily be extended to other machine learning algorithms. Likewise the interface learning technique that was demonstrated to couple multi-fidelity models can be used for coupling multiscale and multiphysics models.

\section*{Acknowledgement}
The authors acknowledge the financial support from the Department of Engineering Norwegian Research Council, the industrial partners of OPWIND: Operational Control for Wind Power Plants (Grant No.: 268044/E20), the U.S. DOE Early Career Research Program (Award Number DE-SC0019290), and the National Science Foundation under Award Number DMS-2012255.


\section*{References}
\bibliographystyle{unsrt} 
\bibliography{ref}

\end{document}